\documentclass[3p,12pt]{elsarticle}
\usepackage{epsfig,graphicx,psfrag,here}
\usepackage{hyperref,natbib,geometry,fleqn,txfonts,endfloat,cleveref}
\newcommand{\bda}{\begin{\displaymath}\begin{array}{rl}}
\newcommand{\eda}{\end{array}\end{displaymath}}
\newcommand{\be}{\begin{equation}}
\newcommand{\ee}{\end{equation}}
\newcommand{\bdm}{\begin{displaymath}}
\newcommand{\edm}{\end{displaymath}}
\newcommand{\bea}{\begin{eqnarray}}
\newcommand{\eea}{\end{eqnarray}}

\newcommand{\indR}{\mbox{\tiny R}}

\newcommand{\QCD}{\mbox{\tiny Q\hspace{-0.05em}CD}}
\newcommand{\QED}{\mbox{\tiny QED}}

\newcommand{\MeV}{\,\mbox{MeV}}
\newcommand{\GeV}{\,\mbox{GeV}}

\newcommand{\al}{&\!\!\!}
\newcommand{\el}{\hspace{0.07em}\mbox{\scriptsize el}}

\begin{document}

\begin{frontmatter}
\title{On the mass difference between proton and neutron}

\author{J.~Gasser}
\ead{gasser@itp.unibe.ch}
\author{H.~Leutwyler}
\ead{leutwyler@itp.unibe.ch}

\address{Albert Einstein Center for Fundamental Physics, Institute for Theoretical Physics, University of Bern, Sidlerstrasse 5, 3012 Bern, Switzerland }

\author{A.~Rusetsky}
\ead{rusetsky@hiskp.uni-bonn.de}

\address{Helmholtz-Institut f\"ur Strahlen- und Kernphysik (Theorie) and Bethe Center for
Theoretical Physics, Universit\"at Bonn, Nussallee 14-16, D-53115 Bonn, Germany}

\address{Tbilisi State University, 0186 Tbilisi, Georgia}

\begin{abstract}
The Cottingham formula expresses the electromagnetic part of the mass of a particle in terms of the virtual Compton scattering amplitude. At large photon momenta, this amplitude is dominated by short distance singularities associated with operators of spin 0 and spin 2. In the difference between proton and neutron, chiral symmetry suppresses the spin 0 term. Although the angular integration removes the spin 2 singularities altogether, the various pieces occurring in the standard decomposition of the Cottingham formula do pick up such contributions. These approach asymptotics extremely slowly because the relevant Wilson coefficients only fall off logarithmically. We rewrite the formula in such a way that the leading spin 2 contributions are avoided ab initio. Using a sum rule that follows from Reggeon dominance, the numerical evaluation of the e.m.~part of the mass difference between proton and neutron yields $m_{\QED}=0.58\pm 0.16\MeV$. The result indicates that the inelastic contributions are small compared to the elastic ones. \end{abstract}

  
\end{frontmatter}

\thispagestyle{empty}
\bigskip
The fact that proton and neutron nearly have the same mass is understood since the 1930s, as consequence of an approximate symmetry: isospin  \cite{Heisenberg}. 
For a long time, it was taken for granted that the symmetry is broken only by the e.m.~interaction -- the Cottingham formula \cite{Cottingham} did explain the size of the observed mass differences with $\Delta I=2$, $m_{\pi^+}-m_{\pi^0}$ for instance.  For $\Delta I = 1$, however, in particular for $m_p-m_n$, the fact that the charged particle is lighter than the neutral one remained mysterious \cite{GrossPagels,ElitzurHarari,Zee}. 

In 1975, Gasser and  Leutwyler  \cite{GL1975} pointed out that the mystery disappears if the popular conviction, according to which the strong interaction conserves isospin, is dismissed. They showed that a coherent picture of isospin breaking can be reached within the Quark Model, provided the masses of the two lightest quarks are not only very small but also very different. At that time, the experimental results on deep inelastic scattering were consistent with the scaling laws of Bjorken \cite{Bjorken}. Evaluating the Cottingham formula with the scarce experimental information available then, they concluded that the elastic contributions dominate over the inelastic ones and arrived at the estimate $m_{\QED}=0.7\pm 0.3\MeV$ \cite{GL1975}. 

Walker-Loud, Carlson and Miller \cite{WCM} claimed that the analysis in \cite{GL1975} is incorrect and instead arrived at $m_{\QED}=1.30(03)(47)$. This paper triggered renewed interest and several authors investigated the matter \cite{ESTY,TWY,Walker-Loud2018,Tomalak}. We will briefly discuss the results obtained in these works below. Some of the claims made in \cite{WCM} are rectified in Appendix E of \cite{GHLR} and in \cite{HLCD15,Hoferichter:2019jhr}. 

A thorough account of our recent work on the subject with the technical details of the calculation is given in \cite{sumrule}. The aim of the present paper is to describe the basic ideas underlying our analysis and the conclusions drawn from it. 

\medskip

 {\it Cottingham formula.}
We denote the e.m.~self-energy of the particle by $m_\gamma$. As shown by Cottingham \cite{Cottingham}, it is determined by the spin averaged forward Compton scattering amplitude, 
\be \label{eq:Compton} T^{\mu\nu}(p,q)=\frac{i}{2}\!\int\!\! d^4x\,e^{i q\cdot x}\langle p|Tj^\mu(x) j^\nu(0)|p\rangle\,.
\ee
Current conservation, Lorentz invariance and symmetry under space reflection imply that $T^{\mu\nu}$  can be expressed 
in terms of two functions $T_1(\nu,q^2), T_2(\nu,q^2)$ that only depend on $\nu=p\cdot q/m$ and $q^2$ ($m$ is the mass of the particle).
We use the notation of \cite{GHLR}:
\bea \label{eq:TK} T^{\mu\nu}(p,q)\al=\al(q^\mu q^\nu-g^{\mu\nu}q^2)T_1(\nu,q^2)\\
  \al+\al \frac{1}{m^2}\{(p^\mu q^\nu+p^\nu q^\mu)p\cdot q
 -g^{\mu\nu}(p\cdot q)^2-p^\mu p^\nu q^2\}T_2(\nu,q^2)\,.\nonumber
\eea
The Cottingham formula represents $m_\gamma$ as an integral over the four components of the photon momentum $q$. In the rest frame of the particle, the analytic properties of the time-ordered product allow one to perform a Wick rotation that turns the path of integration in the variable $q^0$ from the real axis into the imaginary axis, $q^0= i Q_4$ \cite{Cottingham,GHLR}. The variable $\nu$ coincides with $q^0$ and thus becomes purely imaginary. Identifying $Q_1,Q_2,Q_3$ with the space components of the physical momentum,  we have $q^2=-Q^2$, where $Q$ is the length of the euclidean four-vector $Q_\mu $. The Cottingham formula then takes the form of an integral over euclidean space:
\bea\label{eq:euclidean} m_\gamma=\frac{e^2}{2m (2\pi)^4}\int\!\frac{d^4Q}{Q^2}   \phi \,,\quad\quad
\phi =3Q^2 T_1 +(2Q_4^2+Q^2)T_2 \,, \eea
where $T_1,T_2$ are to be evaluated at $\nu=iQ_4$, $q^2=-Q^2$.

\medskip

{\it Operator product expansion.}
The asymptotic beha\-viour of the integrand in formula (\ref{eq:euclidean}) is controlled by the operator product expansion \cite{Wilson,GrossWilczek,Politzer,Weinberg1973,Collins,Buras}. The leading contributions are determined by the Wilson coefficients of the operators of lowest dimension, which carry either spin 0 or spin 2. The explicit expressions \cite{Collins,Buras,HillPaz} show that the contributions from $T_1$ and $T_2$ both diverge -- the formula (\ref{eq:euclidean}) must be regularized, e.g.~by cutting the integral off with $Q^2\leq \Lambda^2$. We denote the regularized version of $m_\gamma$ by $m_\gamma^\Lambda$. 

Since the operators of spin 2 are of anomalous dimension, their contributions to the asymptotic behaviour of $T_1$ and $T_2$ involve a negative fractional power of $\ln Q^2$, so that asymptotics sets in only very slowly. The key observation in our evaluation of the Cottingham formula is that the spin 2 contributions to $T_1$ and $T_2$ are the same -- apart from the sign and a factor of 2.  In the amplitude
\be\label{eq:Tbar} \bar{T}(\nu,q^2)=T_1(\nu,q^2)+\mbox{$\frac{1}{2}$}T_2(\nu,q^2)\,,\ee
the leading short distance singularities of spin 2 drop out. Accordingly, this amplitude approaches asymptotics much more rapidly than the individual terms. The advantage of replacing $T_1$ with $\bar{T}$ also shows up in the asymptotic behaviour of the integrand  in (\ref{eq:euclidean}): the divergence stems from the first term in the decomposition $\phi=3Q^2 \bar{T}-\frac{1}{2}(Q^2-4Q_4^2)T_2$. 
The angular integration suppresses the second when $Q^2\rightarrow\infty$.

\medskip

{\it Dispersion relations}.
 The amplitudes $\bar{T}(\nu,-Q^2)$ and $T_2(\nu,-Q^2)$ obey fixed-$Q^2$ dispersion relations in the variable $\nu$. The imaginary parts are determined by the structure functions: 
\be \mbox{Im}\bar{T}=\pi \bar{F}/2\hspace{0.05em}x \hspace{0.05em}Q^2\,,\quad\quad \mbox{Im}T_2=\pi \hspace{0.1em}2 m^2 x \hspace{0.05em}F_2/Q^4\,,
\quad\quad  \bar{F}=F_L+2\hspace{0.05em}m^2 x^2 F_2/Q^2\,,\ee 
with $x=Q^2/2\hspace{0.05em}m\hspace{0.05em}\nu$ and  $F_L \equiv F_2-2x F_1$. Regge asymptotics implies that $T_2$ obeys an unsubtracted dispersion relation, while $\bar{T}$ requires a subtraction: 
\bea\label{eq:DR} \bar{T}(\nu,-Q^2)\al=\al\bar{T}^{\el}(\nu,-Q^2)+\bar{S} (-Q^2)+
  (Q^2+4\nu^2)\!\!\int_0^{x_{\mbox{\tiny th}}}\hspace{-0.8em}dx\;\frac{m^2 \bar{F}(x,Q^2)}{(Q^2+m^2 x^2)(Q^4-4m^2 x^2\nu^2-i\epsilon)} \,,\nonumber\\ 
T_2(\nu,-Q^2)\al=\al T_2^{\el}(\nu,-Q^2)+\!\!\int_0^{x_{\mbox{\tiny th}}}\hspace{-0.8em}dx\;\frac{4m^2F_2(x,Q^2)}{Q^4-4m^2x^2\nu^2-i\epsilon}\,.
\eea
The elastic parts, $\bar{T}^{\el}$ and $T_2^{\el}$, are unambiguously determined by the nucleon form factors \cite{GL1975,GHLR,Hoferichter:2019jhr}. In the dispersion integrals, we have replaced the variable of integration with $\nu'=Q^2/(2\hspace{0.05em}m\hspace{0.05em}x)$. The upper limit represents the boundary of the inelastic region, $x_{\mbox{\tiny th}}=Q^2/(Q^2+2m M_\pi+M_\pi^2)$. 

Note that we are not subtracting the dispersion integral for $\bar{T}$ at $\nu=0$, but at $\nu=\frac{i}{2}Q$. This simplifies the analysis further, as it implies that, for $\nu=iQ_4$, the subtracted integral picks up the factor $(Q^2-4Q_4^2)$, so that the angular average suppresses that part as well.

\medskip

{\it Renormalization.}
In the framework of QCD+QED, the mass of a particle is determined by the bare parameters that occur in the Lagrangian and the cutoff used to regularize the theory. If the electromagnetic interaction is turned off, only the QCD coupling constant, the quark masses and the cutoff are relevant. To order $e^2$, the e.m. interaction changes the mass not only by the regularized version of equation (\ref{eq:euclidean}), but in addition by the contribution $\Delta m^\Lambda$, which arises from the change in the bare parameters needed for the mass of the particle to stay finite when the cutoff is removed: the bare quantities depend on the cutoff as well as on $e$. 
The e.m.~contribution to the mass is given by
\be m_{\QED}= \mbox{lim}\hspace{-1.8em}\rule[-0.8em]{0em}{0em}_{\Lambda\rightarrow\infty}\,\,\{m_\gamma^\Lambda+\Delta m^\Lambda\}\,.\ee

\medskip

{\it Decomposition of the Cottingham formula.}
With our decomposition of the Compton amplitude, the renormalized Cottingham formula consists of four parts \cite{sumrule}:
 \be\label{eq:decomposition}  m_{\QED}= m_{\el}+m_{\bar{F}}+m_{F_2}+m_{\bar{S}}\,.\ee
While the first term collects the elastic contributions, the second and third ones represent the contributions from the integrals occurring in the dispersion relations for $\bar{T}$ and $T_2$. In the first three parts, the limit $\Lambda\rightarrow\infty$ can be taken -- the explicit expressions involve integrals over the elastic form factors and structure functions of the nucleon \cite{sumrule}. The divergence resides in the fourth term, which contains the contributions from subtraction function and renormalization,
\bea\label{eq:mS}m_{\bar{S}}\al=\al \mbox{lim}\hspace{-1.8em}\rule[-0.8em]{0em}{0em}_{\Lambda\rightarrow\infty}\left\{N\!\!\int_0^{\Lambda^2}\hspace{-1em} dQ^2 Q^2 \bar{S} (-Q^2)+\Delta m^\Lambda\right\}\,,\eea
where the constant $N$ stands for $3\alpha_{em}/8\pi m$.

\medskip

{\it Asymptotic behaviour of the subtraction function.}
The operator product expansion implies that $\bar{S}$ falls off in proportion to $1/Q^4$ when $Q^2$ becomes large , while $\Delta m^\Lambda$ grows logarithmically with $\Lambda$. For $m_{\bar{S}}$ to stay finite, the leading contributions must match:  
\be\label{eq:SC} \bar{S}(-Q^2)\rightarrow\frac{C}{Q^4}\,,\hspace{1em}\Delta m^\Lambda\rightarrow -N C \ln\frac{\Lambda^2}{\mu^2}\,.\ee
The constant $C$ is related to the matrix elements of the lowest dimensional operators of spin 0. 

In the following, we consider the difference between proton and neutron, without explicitly indicating this in the notation: we write $\bar{T}$ for $\bar{T}^{p-n}$ and likewise for $\bar{S}$, $C$, $\bar{F}$, $T_2, \ldots$  In the proton-neutron mass difference, we work to first order in the isospin breaking parameters $m_u-m_d$ and $e^2$ and neglect contributions of $O[e^2(m_u-m_d)]$.  The constant $C$ can then be expressed in terms of the proton matrix elements of $\bar{u}u-\bar{d}d$:
\be\label{eq:C} C= \frac{4m_u-m_d}{9}\langle p|\bar{u}u-\bar{d}d|p\rangle\,.\ee
The same matrix element also determines the leading contribution to the QCD part of the mass difference \cite{GL1982}:
\be\label{eq:mpnQCD}m_{\QCD}=\frac{m_u-m_d}{2m}\langle p|\bar{u}u-\bar{d}d|p\rangle\left\{1+O(m_u-m_d)\right\}\,.\ee
The crude estimate $m_{\QCD}\approx -2\,\mbox{MeV}$ and the known quark mass ratios imply that $C$ is tiny: with the lattice result $m_s/m_{ud}=27.23(10)$ \cite{FLAG} and the value of the ratio $Q\equiv\sqrt{(m_s^2-m_{ud}^2)/(m_d^2-m_u^2)} =22.1(7)$ extracted from $\eta$-decay \cite{CLLP}, we obtain $C\approx 6\cdot 10^{-4} {\GeV}^2$. The approximate chiral symmetry of the Standard Model very strongly suppresses the asymptotic behaviour of $\bar{S}$.

The expression (\ref{eq:C}) for the constant $C$ receives corrections from higher orders of the perturbation series. These imply that the Cottingham formula contains subleading divergences proportional to $\ln\ln \Lambda$ -- for details, we refer to \cite{sumrule}. Since chiral symmetry suppresses the entire contribution from the region where perturbation theory applies, the corresponding effects in the renormalized mass difference are tiny and can be neglected. Setting $\bar{\mu}\equiv\exp(-\frac{1}{2})\mu$, the expression for $\Delta m^\Lambda$ in (\ref{eq:SC}) differs from  $ -N C \int_0^{\Lambda^2} dQ^2 Q^2/(\bar{\mu}^2+Q^2)^2$ only by terms of order $\mu^2/\Lambda^2$. The contribution to $m_{\QED}$ that arises from the subtraction function can thus be written in the form \cite{sumrule}
\be\label{eq:mSbar}m_{\bar{S}}=N\int_0^\infty \hspace{-0.8em}dQ^2 Q^2 \left\{\bar{S}(-Q^2)-\frac{C}{(\bar{\mu}^2+Q^2)^2}\right\}\,.\ee
This distinguishes our evaluation of the Cottingham formula from those in the literature \cite{WCM,ESTY,TWY,Walker-Loud2018,Tomalak}, where the integral as well as the counter term are evaluated at a finite value of the cutoff around $\Lambda^2\approx 2\GeV^2$. This would be legitimate if asymptotics were reached at such a low scale, but for the subtraction function used in these references, that is not the case.

\medskip

{\it Role of spin 2 operators: $\bar{S}$ versus $S_{\hspace{-0.15em}1}$.}
Traditionally, the subtraction function is identified with a multiple of $S_{\hspace{-0.15em}1}(q^2)\equiv T_1(0,q^2)$. Asymptotically, $Q^4 S_{\hspace{-0.15em}1}$ approaches the same constant $C$ as $Q^4\bar{S}$, but this happens much more slowly: in contrast to $\bar{S}$, the function $S_{\hspace{-0.15em}1}$ does pick up spin 2 contributions \cite{Collins} and these are not proportional to the lightest quark masses. This implies that, in the pre-asymptotic region, the two subtraction functions behave quite differently. In $S_{\hspace{-0.15em}1}$, it takes extremely large values of $Q^2$ for the spin 0 term to finally win over those of spin 2.  When working with a low cutoff, this affects the result for the mass difference quite substantially -- see below. 

In contrast to the mass itself,  $m_{\QED}$ depends on the renormalization scale $\mu$: the splitting into a contribution from QCD and one from QED is a matter of convention. In the decomposition (\ref{eq:decomposition}), the parameter $\mu$ resides in $m_{\bar{S}}$ and enters through the term $\Delta m^\Lambda$ in (\ref{eq:SC}). The above estimate for the size of the constant $C$ shows, however, that the sensitivity of  $m_{\bar{S}}$ to $\mu$ is extremely weak: increasing the scale by a factor of 2  increases the value of  $m_{\QED}$ by about 1 keV. For definiteness, we set $\mu=2\,\GeV$.

\medskip

{\it Reggeon dominance.} The behaviour of the Compton amplitude in the limit $q= \lambda\,\bar{q}$, $\lambda\rightarrow\infty$ is controlled by the operator product expansion, which implies that both $\bar{T}$ and $T_2$ tend to zero in this limit. The gluons as well as the quarks reggeize, however \cite{Grisaru1,Grisaru,Lipatov,Kuraev,Balitsky,Gribov2003}: in the limit where $\nu$ becomes large while $q^2$ is kept fixed, only $T_2$ disappears, $\bar{T}$ diverges. 

In the Compton amplitudes of proton and neutron, the leading terms stem from singlet contributions due to Pomeron exchange, but in the difference between the two, these drop out: the dominating contributions to $\bar{T}$ stem from the exchange of the leading Reggeon with $I^C=1^+$, which we refer to as the $a_2$. It generates a Regge pole in the angular momentum plane which moves along the trajectory $\alpha(t)$. In the forward direction, only the value  at $t=0$ matters: at fixed $q^2$, the contribution from the $a_2$ grows with the power $\nu^\alpha$:
\be\label{eq:R} \bar{T}^R(\nu,q^2)=- \frac{\pi\,\beta(q^2)}{\sin\pi\alpha}\{(-s)^{\alpha}+(-u)^{\alpha}\}\,,\ee
where $\alpha$ stands for $\alpha(0)$ and the variables $s =m^2+ 2m \nu+q^2$, $u=m^2-2m \nu+q^2$ represent the square of the centre of mass energy in the $s$- and $u$-channels, respectively. The value of $\alpha$ is experimentally well determined from the high energy behaviour of hadronic cross sections and is in the vicinity of $\alpha\approx 0.55$ -- the uncertainties in $\alpha$ are too small to affect our results.

Reggeization implies that the dispersion relation for $\bar{T}$ requires a subtraction. We assume that the Reggeons fully determine the asymptotic behaviour \cite{ElitzurHarari,GL1975,GHLR},
\be\label{eq:RD}\lim\rule[-0.7em]{0em}{1em}_{\hspace{-1.7em}\nu\rightarrow\infty} \,(\bar{T}-\bar{T}^{\indR})= 0\,,\ee
and that the remainder disappears sufficiently fast for the difference $\bar{T}-\bar{T}^{\indR}$ to obey an unsubtracted dispersion relation. We refer to this assumption as Reggeon dominance. 

A nonzero limiting value in (\ref{eq:RD}) would represent a fixed pole in $\bar{T}$ at $\alpha=0$. We do not know of a physical phenomenon that could produce such a term -- neither causality, nor the short-distance singularities, nor the Reggeons generate terms of this sort. The presence of a fixed pole would mean that the high energy behaviour of the Compton amplitude is not understood. 

For $T_2$,  the contribution from the Reggeons tends to zero in proportion to $\nu^{\alpha-2}$. The generally accepted assumption that this amplitude obeys an unsubtracted dispersion relation immediately implies that it also obeys the Reggeon dominance condition (\ref{eq:RD}).  Note, however, that the expansion of the dispersion integral for $T_2$ in inverse powers of $\nu$ contains a term proportional to $1/\nu^2$. As pointed out by Damashek and Gilman \cite{DG} and, independently by Dominguez et al.~\cite{Dominguez}, this term corresponds to a fixed pole in $T_2$, at $\alpha=0$ (there is an analogous term also in $\bar{T}$, but it represents a fixed pole with $\alpha=-2$ and is at most of academic interest).  

\medskip

{\it Sum rule.}
Elitzur and Harari \cite{ElitzurHarari} pointed out that if the exchange of Reggeons correctly describes the asymptotic behaviour in the limit $\nu\rightarrow \infty$ at fixed $q^2$, then the subtraction function obeys a sum rule which fully determines it through the cross section of lepton-nucleon scattering.  The sum rule relevant for our decomposition of the Compton amplitude exclusively involves the structure function $\bar{F}$.  At small values of $x$, this quantity is dominated by Reggeon exchange:
\be\label{eq:FR}\bar{F}^R=  b (Q^2)x^{1-\alpha}\,,\hspace{1em}b(Q^2)=2Q^{2(\alpha+1)}\beta(-Q^2)\,.\ee 
In the difference $\bar{F}-\bar{F}^R$, the leading term cancels out. As demonstrated in \cite{sumrule}, the sum rule for $\bar{S}$ can be brought to the form:
\bea\label{eq:Sum rule}Q^2\bar{S}(-Q^2)=\int_0^{x_{\mbox{\tiny th}}}\hspace{-0.8em}
dx\;\frac{\bar{F}(x,Q^2)-\bar{F}^R(x,Q^2)}{x^2}- \frac{b(Q^2)}
{\alpha\, x_{\mbox{\tiny th}}^\alpha}-\int_0^{x_{\mbox{\tiny th}}}\hspace{-0.8em}
dx\; \frac{m^2\bar{F}(x,Q^2)}{Q^2+m^2 x^2}\,.\eea
 
In \cite{GL1975}, the violations of Bjorken scaling were ignored: it was assumed that for $Q^2\rightarrow\infty$,  the structure function $\bar{F}$ tends to $(2xH_1+F_2)x^2m^2/Q^2$, where $H_1$ and $F_2$ only depend on $x$.  One readily checks that the sum rule (\ref{eq:Sum rule}) then indeed reduces to the relation between the operator matrix element $C$ and the structure functions given in (5.2), (5.3), (13.14) of \cite{GL1975}. Scaling would imply that the last term on the r.h.s.~of (\ref{eq:Sum rule}) tends to zero $\propto1/Q^4$. The scaling violations merely make it disappear less rapidly, in proportion to $1/Q^2/(\ln  Q^2)^{ 1+d_2}$ with $d_2>0$ \cite{sumrule}.

\medskip

{\it Elastic contributions.}

In recent years, the precision to which the elastic form factors are known has increased significantly \cite{Formfactors,Kelly,Ye2018,Borah}. The results obtained with the  three parametrizations of \cite{Formfactors,Kelly,Ye2018} are in the range 
\be \label{eq:mel}m_{\el}=0.75\pm 0.02\MeV\,.\ee

Note that the amplitudes used in the literature often have kinematic zeros -- this can make it difficult not only to sort out the asymptotic behaviour, but also to identify the elastic part of the dispersive representation (``Born term'') with the contribution generated by the one-particle intermediate state \cite{GHLR,Hoferichter:2019jhr}. In \cite{Tomalak}, for instance, it is assumed that the amplitude $\hat{T}=q^2T_1+\nu^2 T_2$ satisfies the asymptotic condition (\ref{eq:RD}). That assumption, however, requires $q^2T_1$ to contain a fixed pole which compensates the one in $\nu^2 T_2$ and hence violates Reggeon dominance.

\medskip

{\it Structure functions, parton distributions.}
For the numerical evaluation of the dispersion integrals and of the sum rule for the subtraction function, we need a representation for the difference between the structure functions of proton and neutron, and not only for the relatively well explored quantity $F_2$, but also for the longitudinal component $F_L$, which is known less well. In the resonance region ($W<3$), we make use of the parametrizations of the structure functions in  \cite{Drechsel:2007if,Kamalov:2000en,Hilt:2013fda,Bosted1,Bosted2}. For $W>3$ and low photon virtualities ($Q^2<1$), we invoke the Regge representation of Alwall and Ingelman (AI) \cite{GVMD}. 

At higher values of $Q^2$, the DGLAP equations \cite{DGLAP,DGLAP1,DGLAP2} for the parton distribution functions (PDFs)
 provide a strong constraint for the analysis of the data: at leading order in $\alpha_s$, these equations imply that $F_L$ is given by an integral over $F_2$. A vast amount of PDFs is available \cite{Andersen:2014efa,Buckley:2014ana} and APFEL Web \cite{Bertone:2013vaa,Carrazza:2014gfa} provides a flexible, user-friendly tool for the evaluation
of the corresponding structure functions. Since the quark masses $m_u$ and $m_d$ are tiny, the $u$- and $d$-distributions in the neutron must be very close to the $d$- and $u$-distributions in the proton, respectively. As emphasized e.g.in \cite{ABM2}, this ensures  that the $u$- and $d$-distributions can separately be determined by using neutral and charged current data on the proton -- scattering on deuterons or heavier nuclei is not needed to sort out the difference between the $u$- and $d$-distributions of the proton. 

As mentioned above, the behaviour in the Regge region ($x$ small) is dominated by the singlet part of the distributions. This implies that the $u$- and $d$-distributions must approach one another when $x\rightarrow 0$.  While the available data strongly constrain the singlet part at small $x$, the non-singlet PDFs are much less well determined. The same applies to the non-singlet structure function $\bar{F}$ which plays a central role in our work. 

Since reggeization involves sums to all orders of perturbation theory, it is not a simple matter to analyze the behaviour at small $x$ in the framework of the DGLAP equations (for a review of the problems encountered in this endeavour, we refer to \cite{Cooper-Sarkar}). 
In particular, the requirement that the $u$- and $d$-distributions must approach one another in the Regge limit and that their difference yields a contribution to $\bar{F}$ that falls off with $b(Q^2) x^{1-\alpha}$, $\alpha\approx 0.55$ imposes nontrivial theoretical constraints. A coherent parametrization of the PDFs that is consistent not only with the data, but also with these constraints, yet needs to be found. 

In our calculation, we rely on the solutions of the DGLAP equations constructed by Alekhin, Bl\"umlein and Moch (ABM) \cite{ABM2,ABM1,ABM3} in the region $x>0.01$. At smaller values of $x$, we assume that $\bar{F}$ is dominated by the Reggeon $a_2$, not only at small photon virtualities, but also at higher values of $Q^2$. We determine the residue $b(Q^2)$ by smoothly matching the two parametrizations around $x=0.01$. In the  region $W>3$, $Q^2>1$, we estimate the uncertainty in our representation for the structure functions  $\bar{F}\rule[0.02em]{0em}{0.75em}^{\hspace{0.03cm}p-n}$ and $F_2^{p-n}$ at 30\%. For details, we refer to \cite{sumrule}. 

\medskip

{\it Inelastic contributions.}  
The most striking aspect of our numerical result is that the two terms $m_{\bar{F}}$ and $m_{F_2}$ turn out to be tiny: 
$m_{\bar{F}}+m_{F_2}= -0.004(1)\MeV$.
As discussed above, the angular integration suppresses the integrands of these quantities at large values of $Q^2$, but the numerical result shows that the suppression is very efficient also in the low energy region. 
We conclude that -- in our decomposition of the Compton amplitude -- only the elastic contribution $m_{\el}$ and the term $m_{\bar{S}}$ from the subtraction function play a significant role.  Note that this statement holds independently of the assumptions used to determine the subtraction function. Since the various attempts at evaluating the Cottingham formula arrive at very similar values for the elastic part, the discrepancies in the results for the mass difference mainly come from the term $m_{\bar{S}}$, i.e. from the fact that the parametrizations used for the Compton amplitude yield different representations for the subtraction function $\bar{S}$.  

\medskip

{\it Contribution from the subtraction function.} It is straightforward to evaluate the sum rule for $\bar{S}$ with the two representations of $\bar{F}(x,Q^2)$ and $b(Q^2)$ we are using below and above $Q^2=1\GeV^2$, respectively and to calculate the corresponding contribution to $m_{\bar{S}}$ with (\ref{eq:mSbar}). Isospin conservation prevents the most prominent feature in the low energy region, the $\Delta(1232)$, to make a significant contribution. Moreover, the regions below and above a centre of mass energy of $3\GeV$ contribute with opposite sign -- within errors, they cancel: $m_{\bar{S}}(Q^2\!<\!1\GeV^2)=-0.034(68)\MeV$. Note that the error is twice as large as the central value. It is dominated by the uncertainties in the resonance region and is of systematic nature, as it stems from the simplification used in the data analysis of Bosted and Christy \cite{Bosted1,Bosted2}: the ratio $R=\sigma_L/\sigma_T$ is assumed to be the same for proton and neutron. In the region where the Pomeron dominates, this holds to good accuracy, but we need the difference between the two, where Pomeron exchange drops out.   

At $Q^2=1\GeV^2$, where the representations AI and ABM meet, the results for the contributions to $\bar{S}$ from $W>3\GeV$ agree within errors:  the two entirely different sources match, both in sign and in size.  
In order to interpolate between the values of $Q^2$ where the ABM results provide significant information and the region where asymptotics sets in, we make use of the Generalized Vector Dominance Model of Sakurai and Schildknecht \cite{SakuraiSchildknecht}, parametrizing the subtraction function in terms of the contributions from $\rho$, $\omega$ and $\phi$. In the difference between proton and neutron, only the off-diagonal terms survive:
\be\label{eq:SVMD} \bar{S}_{\mbox{\tiny VMD}}(-Q^2)=\frac{1}{m_\rho^2+Q^2}\left\{\frac{c_\omega}{m_\omega^2+Q^2}+\frac{c_\phi}{m_\phi^2+Q^2}\right\}\,.\ee
The asymptotic condition (\ref{eq:SC}) requires the two terms in the bracket to nearly cancel: $c_\omega+c_\phi=C$. This leaves a single parameter free, say $c_\omega$. Fitting the parametrization in the range between 2 and $3.5\GeV^2$, we obtain
$ c_\omega=-0.74(49)\GeV^2$. We have checked that the outcome for $m_{\bar{S}}$ is neither sensitive to the specific form of the interpolation  nor to the range used in the fit. Numerically, this yields $m_{\bar{S}}(Q^2\!>\!1\GeV^2)=-0.13(9)\MeV$. Together with the contributions from low virtualities, this yields
\be m_{\bar{S}}=-0.17(16)\MeV\,.\ee

Because asymptotic freedom fixes the asymptotic behaviour of the subtraction function, the parametrization obtained within Generalized Vector Meson Dominance contains a single free parameter. Instead of fixing it to the results obtained with the ABM solution of the DGLAP equations, we can dismiss the experimental information available for $W>3$, $Q^2>1$ altogether and determine the free parameter with a fit to the results obtained for $Q^2<1$. This yields $m_{\bar{S}}=-0.12(21) \MeV$: the central value stays well within the estimated uncertainty and the error only increases by about 30\%. This indicates that our result is not sensitive to the input used in the region where the non-singlet contributions to the structure functions are not yet known well. 

\medskip

{\it Numerical result.}
Collecting the various contributions and using the experimental value of the proton-neutron mass difference, the parts due to the e.m.~interaction and to the difference between $m_u$ and $m_d$ become  
\be\label{eq:meQED} m_{\QED}= 0.58\pm 0.16\MeV\,,\hspace{0.3em}m_{\QCD}=-1.87\mp 0.16\MeV\,.\ee 
The result for $m_{\QCD}$ yields a more precise estimate for the leading Wilson coefficient: $C= 5.7(1.1)\cdot 10^{-4}\GeV^2$, but the corresponding shift in our results is negligibly small. 

The conclusions reached in \cite{GL1975} are thus confirmed: $m_{\QED}$ is dominated by the elastic contribution. The uncertainty in the old result, $m_{\QED}= 0.7(3)\MeV$, is reduced by about a factor of two. 

It is not difficult to understand why the inelastic contributions are so small: (a) the angular integration suppresses the contributions from the dispersion integrals, (b) if $Q^2$ is large, the subtraction functions of proton and neutron are nearly the same -- in the chiral limit, there is no difference, (c) in the region where Reggeon exchange dominates, the leading term, the Pomeron, is the same, (d) isospin symmetry ensures that the most important resonance, the $\Delta(1232)$, contributes equally to proton and neutron and (e) the elastic contributions also dominate the chiral perturbation series of the Compton amplitude: the leading terms exclusively contribute to the elastic part - inelastic processes merely generate higher order corrections (for a recent analysis of the subtraction function in $\chi$PT, we refer to \cite{Lozano}). 

\medskip

{\it Comparison with lattice results.}
The determination of $m_{\QED}$ on a lattice is a very demanding goal. While the numbers in \cite{Endres:2015gda} cluster around $m_{\QED}\approx 0.7 \MeV$, in agreement with our result, the values $1.00(7)(14)\MeV$ \cite{Borsanyi2015}, $1.03(17)$ \cite{Brantley:2016our} and $1.53(25)(50)\MeV$ \cite{Horsley:2019wha} are higher than ours. Adding statistical and systematic errors in quadrature, the various lattice results differ from the outcome of our calculation by less than two standard deviations. 
In the framework we are relying on, values like $m_{\QED}=1\MeV$ or even higher require sizeable positive contributions from $m_{\bar{S}}$  -- this is not compatible with Reggeon dominance. 

\medskip

{\it Comparison with other dispersive calculations.}
The main difference between our analysis and the work reported in \cite{WCM,ESTY,TWY,Walker-Loud2018}  is that, there, the subtraction function is not calculated, but parametrized with an ansatz in terms of its value at $Q^2=0$ (taken from experiment) and a scale $m_0$ that specifies the momentum dependence.  Moreover, in these models, the parametrization is applied to $S_{\hspace{-0.15em}1}$ rather than to $\bar{S}$. As discussed above, the asymptotic behaviour of $S_{\hspace{-0.15em}1}$ picks up contributions from operators with spin 2, which fall off only extremely slowly. 
Since chiral symmetry suppresses the coefficient $C$ of the leading asymptotic term, it starts dominating $S_{\hspace{-0.15em}1}$ only if $Q^2$ becomes very large. This implies that the parametrization used in these models does not behave properly in the pre-asymptotic region, which does make a significant contribution to $m_{\bar{S}}$. 

The mismatch with the asymptotics disappears if the ansatz in \cite{ESTY} is assumed to be valid for $\bar{S}$ rather than $S_{\hspace{-0.15em}1}$. The central value obtained for $m_{\QED}$ then drops to about $0.7\MeV$, in agreement with what we find. The uncertainties in the result for $m_ {\QED} $, however, are much larger than ours, not only because the experimental values of the magnetic polarizabilities of proton and neutron, which play a key role in those models, are subject to large errors, but also because the result is quite sensitive to the shape of the parametrization used for low values of $Q^2$.

\medskip

{\it Summary.} We have applied the Reggeon dominance hypothesis to the electromagnetic part of the
proton-neutron mass difference.
The uncertainty in our final result, $m_{\QED}=0.58\pm 0.16\,\mbox{MeV}$,
stems from a careful estimate of the errors coming from the 
different experimental data sets used in the calculations. While this confirms the old result \cite{GL1975}, which also relies on Reggeon dominance, recent evaluations of  the Cottingham formula \cite{WCM,ESTY,TWY,Walker-Loud2018,Tomalak}
yield central values around $1\MeV$ or even higher. The difference stems from 
the short distance singularities associated with operators of spin 2 that are
neglected in those references. The lattice determinations do not yet yield conclusive
results, but the method is gradually improved. In the long run, these will achieve comparable 
accuracy and thereby put Reggeon dominance to a very stringent test.  

\section*{Acknowledgements}
 
 We thank Johannes Bl\"umlein for providing us with numerical tables for the structure functions based on the ABM solutions of the DGLAP equations, Thomas Becher for information about the anomalous dimensions in QCD+QED, Vadim Lensky and Vladimir Pascalutsa for a Mathematica notebook concerning the representation of the Compton amplitude in $\chi$PT and Ryan Bignell, Irinel Caprini, Stefano Carrazza, Gilberto Colangelo, Cesareo Dominguez, Franziska Hagelstein, Bastian Kubis, Ulf-G.~Mei{\ss}ner, Sven-Olaf Moch, Gerrit Schierholz and Ignazio Scimemi for comments and useful information. 
A.R. acknowledges the support from the DFG (CRC 110 ``Symmetries 
and the Emergence of Structure in QCD''), as well as from
Volkswagenstiftung under contract no. 93562 and the Chinese Academy
of Sciences (CAS) President's International Fellowship Initiative (PIFI) (grant no. 2021VMB0007).

\biboptions{sort&compress}
\end{document}